# X-Shooting EF Eridani: further evidence for a massive white dwarf and a sub-stellar secondary ⋆

A.D. Schwope[1] and L. Christensen[2,3]

[1] Astrophysikalisches Institut Potsdam, An der Sternwarte 16, 14482 Potsdam, Germany e-mail: `aschwope@aip.de`
[2] European Southern Observatory, Casilla 19001, Santiago 19, Chile
[3] Excellence Cluster Universe, Technische Universität München, Boltzmanstrasse 2, 85748 Garching, Germany

**ABSTRACT**

High resolution spectral observations of the polar EF Eridani obtained in a low state with X-Shooter revealed narrow emission lines from the irradiated secondary. The lines were most prominent in the near-IR Ca ɪɪ-triplet, the more intensive H$\alpha$ line had additional emission likely originating from an accretion stream. The lines with a radial velocity amplitude, $K'_2 = 385 \pm 4 \,\mathrm{km\,s^{-1}}$, serve as tracer of the otherwise unseen companion. The mass function implies a massive white dwarf with $M_\mathrm{wd} > 0.65\,M_\odot$ at $3\sigma$ confidence, and a short distance to the binary, $d \sim 111$ pc ($< 145$ pc at $3\sigma$ confidence). The spectral energy distribution from the UV to the IR together with the high mass ratio gives further strong evidence of EF Eri being a post period-minimum object with $M_2 < 0.06\,M_\odot$.

## 1. Introduction

EF Eridani was only the third Polar (AM Herculis binary) discovered after AM Herculis itself and VV Puppis and the first one discovered initially as a bright X-ray source with spectroscopic and polarimetric identification afterwards (Williams et al. 1979; Griffiths et al. 1979; Tapia 1979). The long-term light curve assembled from the Harvard plate stacks displays variability of the binary between $B = 15 - 18$ mag (Griffiths et al. 1979), naturally explained as high and low accretion states. Since its discovery in 1978 it remained in a high state untill 1997 (Wheatley & Ramsay 1998); the current low state was interrupted by just three short-lasting accretion events (Mukai et al. 2008).

During high and low states EF Eri behaved as the textbook example of AM Herculis binaries (Beuermann et al. 1987; Schwope et al. 2007), at least as far as energy conversion and release in the accretion column was concerned. It changed from shock-heating/bremsstrahlung cooling in the high state to particle heating/cyclotron cooling in the low state, when the overall accretion rate is reduced by a factor 100 or more.

The high state emission line spectrum of EF Eri was studied by Verbunt et al. (1980); Bailey & Ward (1981); Hutchings et al. (1982); Mukai & Charles (1985). They describe a complex multi-component line pattern of high and low ionization species, nevertheless stable from epoch to epoch. None of the several line components could be uniquely associated with a physical structure in the binary apart from the high-velocity line wings which clearly originate in streaming matter between the two stars.

The low state gave view to the component stars in EF Eri. A complex magnetic field topology for the white dwarf with strong quadrupolar components was infered from Zeeman tomography based on low-resolution VLT/FORS spectra (Beuermann et al. 2007). Evidence for a high-mass white dwarf, $M_\mathrm{wd} = 0.82 - 1.00\,M_\odot$ was derived by fitting the GINGA X-ray spectrum (Cropper et al. 1998). Since the model of Cropper et al. in some cases and for unknown reasons revealed masses in stark contrast to mass estimates based on radial velocities or spectroscopic parallaxes (see e.g. Gänsicke et al. 1995; Catalán et al. 1999, for AM Her and QQ Vul, respectively) independent confirmation is necessary. Recently Howell et al. (2006) presented spectroscopic monitoring of EF Eri with the 1.3m SMARTS telescope and they report the detection of several lines in the low state which they ascribe to chromospheric activity of the secondary. H$\alpha$ was persistently observed throughout the orbital cycle. A sine fit to the H$\alpha$ radial velocity curve revealed $K_2 = 269 \pm 18\,\mathrm{km\,s^{-1}}$, regarded by them as projected orbital velocity of the secondary.

EF Eri is a key object for binary stellar evolution with high-field white dwarfs. Close to the CV period minimum, $P_\mathrm{orb} = 81$ min, it may harbour a late-type main-sequence star or, as a period bouncing system, a brown dwarf-like degenerate object as secondary star. Evidence for a sub-stellar secondary was derived from an upper limit TiO band strength in the first low-state NTT spectra (Beuermann et al. 2000). Likely spectroscopic and photometric signatures were reported by Howell & Ciardi (2001) and Harrison et al. (2003) but the claims of the detected secondary remained unconfirmed. Infrared cyclotron emission from a tenuous accretion plasma is definitely contributing to the excess infrared emission, the contribution of the low mass donor is still uncertain (Harrison et al. 2004; Schwope et al. 2007; Campbell et al. 2008).

⋆ Based on observations made with ESO Telescopes at the Paranal Observatory under programme ID 60.A-9436(A)

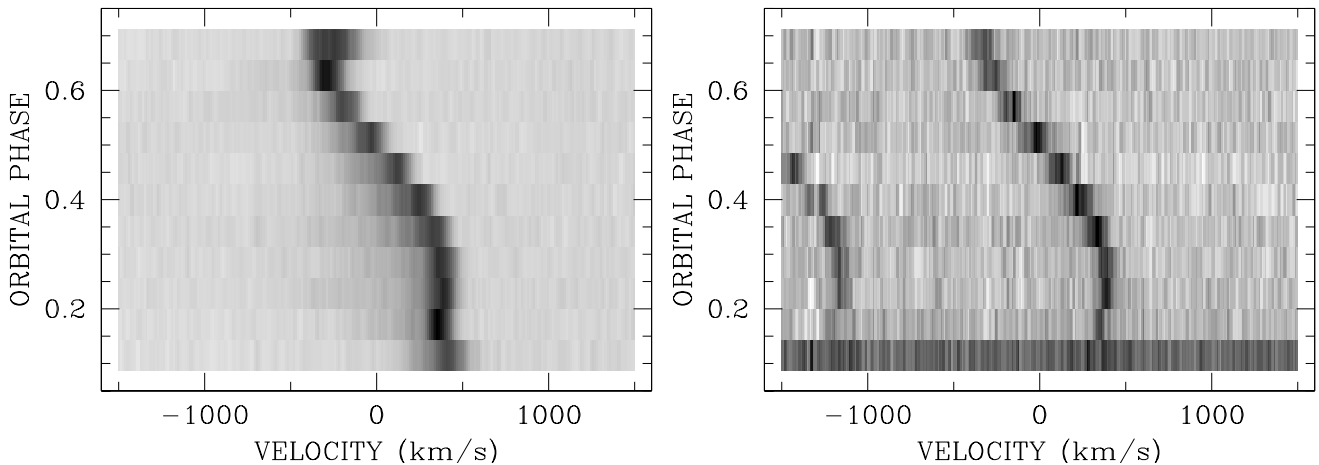

**Fig. 1.** Trailed spectrograms of EF Eri of the Hα (left) and Ca II 8542 emission lines (right) obtained August 13, 2009, in a low state. The additional line in the right panel is Ca II 8498.

To search for lines useful for a radial velocity study and establish the spectral energy distribution from the ultraviolet to the infrared we proposed to include EF Eri in the Science Verification programme for the VLT 2nd generation instrument X-Shooter (D'Odorico et al. 2006). The observations took place on August 13, 2009, and the raw data were made available on August 25. We take the opportunity to single out one important result from an early pipeline reduction of the complex data set, highlighting the capability of the instrument for binary star research and addressing an important issue in EF Eri, its radial velocity solution, and the mass constraints which rest on it.

## 2. X-Shooter spectroscopy of EF Eri

EF Eri was selected for X-Shooter Science Verification observations for a total of $1^h$. One observation block was created to be executed in service mode. Telescope and instrument overheads reduced the time on target to 2080 s, which was split in equal parts to obtain 11 spectral exposures of length 189 s each. The observations started August 13, 2009, at UT 07:45:32, and covered binary cycle 197488.508 − 197489.076 using the updated ephemeris of the X-ray dip Beuermann et al. (2007) and the period derived by Piirola et al. (1987). A phase resolution of 0.056 of the $P_{orb} = 4861$ s binary was achieved.

The seeing was slightly variable between $0\rlap{.}''83$ and $1\rlap{.}''02$ thus always almost matching the chosen slit width of $1\rlap{.}''2$. Spectrum #7 at phase 0.46 was obtained under the least optimum seeing, the continuum and line flux suffer slightly from likely slit loss (see Figs. 1 and 3).

The X-shooter data were reduced with a pre-release version of the pipeline (Goldoni et al. 2006). Data from the three arms were reduced with similar procedures. The pipeline uses a standard bias subtraction in the UVB and VIS arms, and dark frame subtraction in the NIR arm. Sky emission lines are subtracted from the curved orders using a resampling procedure (Kelson 2003). The orders are traced, extracted and wavelength calibrated and rectified. The final results provide frames of both individually extracted orders and their associated errors, and a frame with all the orders merged weighted according to the errors.

One-dimensional spectra were finally extracted with standard MIDAS procedures. The spectra were binned on a 0.25 Å grid in the visual arm, the spectral resolution was estimated from the FWHM width of night sky lines to $R = 6000$ at 8500 Å, typical uncertainties of the wavelength calibration were determined to be $0.1 − 0.2$ pixels. No science-grade spectral response function for our observing night was available at the time of writing. However, a quantitative analysis of the main emission lines is nevertheless possible currently and we concentrate in this paper on the spectral region around Hα and the Ca II-triplett which are covered by spectral orders 25 and 19 of X-Shooter's visual arm, respectively.

Continuum flux variability over the 45 min of observations was found to be below 7% at short wavelengths, < 7200 Å. This low level could already be caused by slit losses due to variable seeing and is not necessarily intrinsic to the source. A red spectral component longward of ∼7200 Å likely similar to that observed by Wheatley & Ramsay (1998) was present in the first 2 spectra. Pronounced Zeeman absorption lines from the white dwarf and the low level of orbital variability clearly reveal that EF Eri was observed in a low state of accretion.

The 11 spectra were arranged in a two-dimensional image to mimic a trailed spectrogram. A few emission lines could be recognized, most prominently Hα, HeI5875 and the Ca II-triplet (8498, 8542, 8662). An average spectrum around the Ca II-triplet is shown in Fig. 2 and cutouts of the trailed spectrogram around Hα and the most prominent Ca II line at 8542 Å are displayed in Fig. 1.

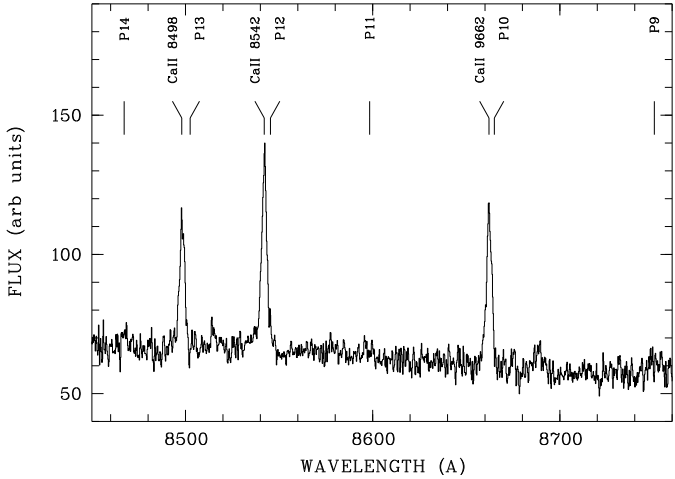

**Fig. 2.** Mean X-Shooter spectrum of EF Eri (visual arm, 19th order) of exposures #3 – #9. The individual spectra were transformed to zero radial velocity before averaging. Observed Ca<small>II</small> and possible H-Paschen lines are indicated.

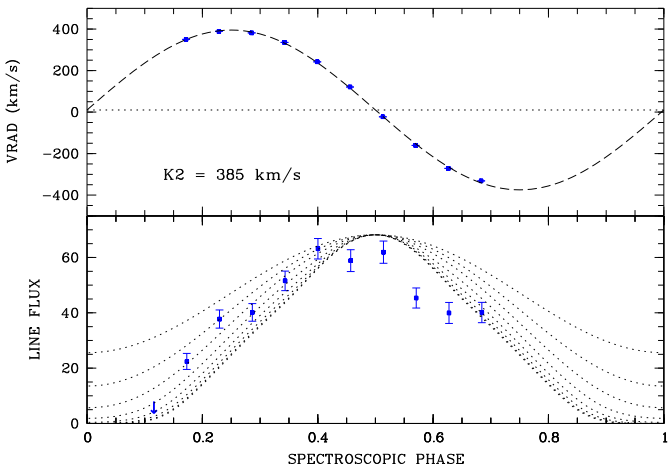

**Fig. 3.** Radial velocity and flux variations of the Ca<small>II</small> 8542 emission line. The dashed line in the upper panel indicates the best fitting sine curve and dashed lines in the lower panel give predicted line flux variations based on our irradiation model for orbital inclinations $i = 20°, 30° \ldots 90°$ (from top to bottom).

While H$\alpha$ has a complex line profile with a sharp component varying sinusoidally and line wings extending to several hundred kilometers away from the narrow core, the Ca<small>II</small> lines display just the narrow line component. It displays pronounced flux variability with maximum flux occurring around orbital phase 0.5 (phase convention derived below). Giving the striking similarity to line profile variability of other polars observed in high and low states (see the collection of Doppler maps in e.g. Schwope 2001, and references therein) the narrow line component can unequivocally be assigned to the secondary star. The H$\alpha$ line has additional components, perhaps best visible in the first spectrum at phase 0.12. Another rather narrow component offset from the sine curve of the narrow emisison line can be recognized therein.

We measured radial velocities of the Ca<small>II</small> 8542 line by fitting single Gaussians to the observed line profiles. We focus here on the brightest of the three lines for signal/noise considerations. A line position could not be measured in the first spectrum due to the faintness of the line. In Fig. 2 the positions of the Ca<small>II</small> lines and of the Hydrogen Paschen lines are indicated. The radial velocity corrected average spectrum displays weak Paschen emission lines of P9, P11, and P14, respectively. Hence, the blending lines P10, P12, and P13 are weakly contributing to the Ca<small>II</small> lines but due to their faintness cannot be recognized in the individual spectra and are not considered in our analysis.

Results of radial velocity measurements of the Ca<small>II</small>8542 line are shown in Fig. 3. An unweighted sine fit to the radial velocity curve revealed a radial velocity amplitude $K'_2 = 385 \pm 4$ km s$^{-1}$. The epoch of blue-to-red zero crossing at BJD = 2455056$.^{\rm d}$8188 $\pm$ 0$.^{\rm d}$0003 determines orbital phase $\phi = 0.0$ throughout this paper and is a robust measurement of inferior conjunction of the secondary in EF Eri. Deviations from the sine curve are negligible. The line flux is highly variable, reaching a maximum plateau between phases 0.4 and 0.55 and declines before and thereafter. Similarly variable is the FWHM line width which is ~130 km s$^{-1}$ in the phase interval $\phi = 0.4 - 0.65$ (i.e. the lines are resolved in this phase interval) and gradually declines down to the instrumental resolution of ~50 km s$^{-1}$ observed in spectrum #2 at phase 0.17.

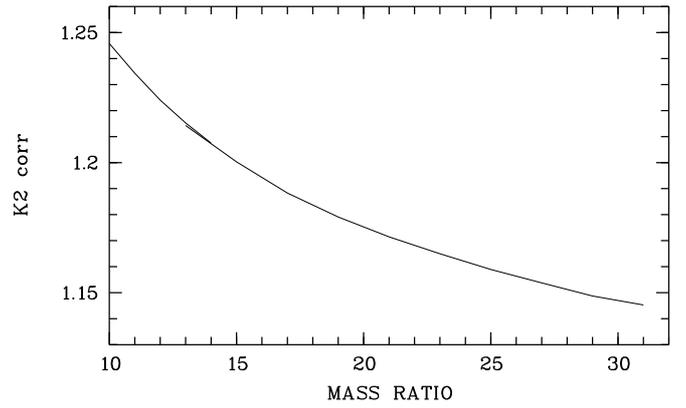

**Fig. 4.** Radial velocity factor $K_2/K'_2$ to correct from center of light to center of mass based on our irradiation model for a star filling its Roche lobe.

### 2.1. Results and analysis

The line profile variations of the Ca<small>II</small> lines leave little doubt about their origin. They were emitted from the hemisphere of the secondary facing the white dwarf and being irradiated by X-ray and UV radiation from the accretion spot and white dwarf. This will be our working hypothesis in the following.

To constrain the mass of the white dwarf we need to estimate the true orbital velocity (corrected for irradiation and inclination) and the mass of the secondary.

We use an irradiation model (Beuermann & Thomas 1990) to correct the observed $K'_2$ from center-of-light to center-of-mass and by that predict the true orbital velocity. The resulting correction factor is a function of the mass ratio in the first place. The very weak dependance on the orbital inclination is negligible. The correction factor is shown in Fig. 4, it is for all acceptable mass ratios larger than 15%.

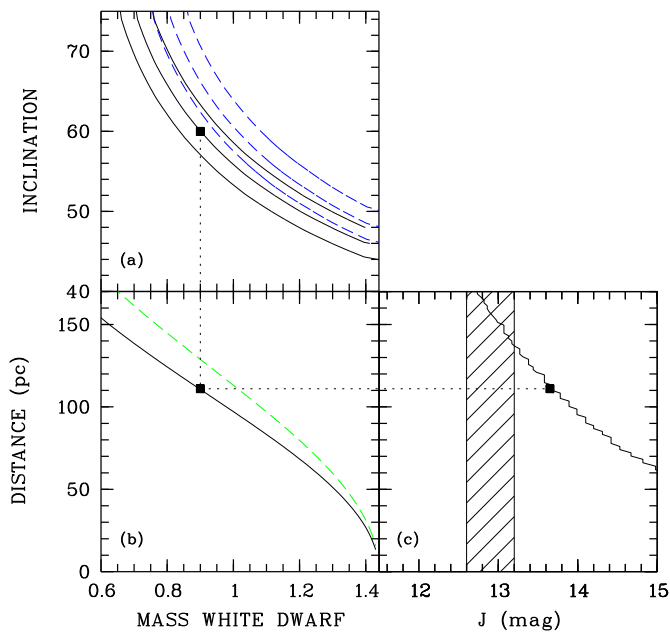

**Fig. 5.** (a) White dwarf mass function based on our irradiation model. Solid lines are for an $0.045\,M_\odot$, dotted lines for an $0.072\,M_\odot$ secondary. The three lines each illustrate the $3\sigma$ range around the nominal velocity $K'_2 = 385$ km s$^{-1}$. (b) White dwarf distance in EF Eri for assumed temperatures of 10000 K (black solid line) and 11000],K (green dashed line) (c) Maximum J-band absolute brightness of the secondary in EF Eri for the given distance. The black solid line was computed for a J-band flux from the secondary of $0.09 \times 10^{-16}$ erg cm$^{-2}$ s$^{-1}$ Å$^{-1}$, the dashed red line for $0.18\times10^{-16}$ erg cm$^{-2}$ s$^{-1}$ Å$^{-1}$. Within the shaded region the transition from a stellar (left) to a sub-stellar (right) secondary (left) occurs according to Knigge (2007).

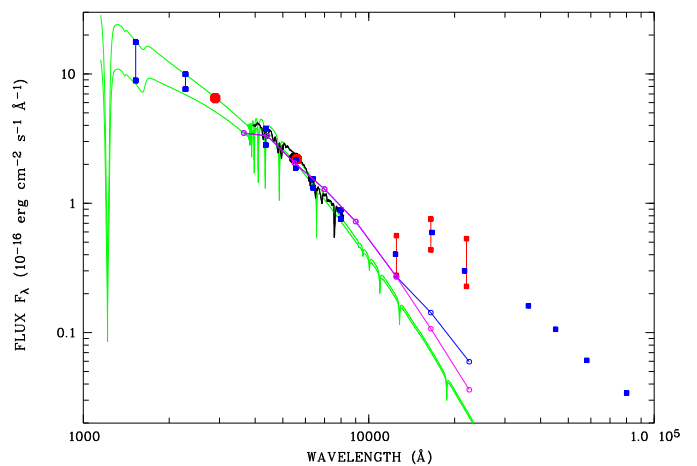

**Fig. 6.** Spectral energy distribution of EF Eri. For an explanation of the used symbols and the data displayed see text.

The orbital inclination of EF Eri was constrained by the occurrence of X-ray absorption dips to $i > \delta$. The angle $\delta$ is defined by the local magnetic field vector in the accretion spot on the surface of the white dwarf (see e.g. Beuermann et al. 1987). The inclination was further constrained by cyclotron models applied to photoelectric polarimetry (Cropper 1995; Piirola et al. 1987), and by modeling X-ray and UV light curves (Beuermann et al. 1987; Schwope et al. 2007). All authors agree that the inclination is most probably between 50° and 65°, although values as high as $i = 75°$ cannot be completely excluded.

The inclination influences the visibility of the irradiated region of the secondary and thus gives an imprint on the line flux variability. It can be estimated from the present observations. We use the irradiation model to calculate Doppler-broadened line profiles and generate emission-line light curves as a function of the orbital inclination. Results are shown in Fig. 3 with a normalization chosen by eye, the same value for all light curves. Line flux variability rules out $i < 40°$ in agreement with previous estimates. The phase coverage and the signal-to-noise ratio of the present data is insufficient to draw more stringent constraints on $i$. There are differences between the model and the data around maximum phase, e.g. the models do not show the plateau between $\phi = 0.4 - 0.6$. The slight depression of the line flux in spectrum #7 at phase 0.46 is likely due to slit losses and thus apparent only.

The mass of the secondary could not too strongly constrained observationally in the past, $M_2 < 0.142$ (see the discussion in Beuermann et al. 2000). We thus make use of Knigge's donor sequence (Knigge 2006, 2007) to further constrain the likely mass of the white dwarf. At the orbital period of 4861 s close to the CV period minimum the object might be a pre-period minimum ordinary CV or a post period minimum object, a so-called period bouncer. Knigge's relation predicts $M_2 = 0.072\,M_\odot$ for the first, $M_2 = 0.045\,M_\odot$ for the second case. Using the irradiation model we calculated $K'_2$ as a function of the mass ratio and of the inclination for the two assumed masses. The curves in Fig. 5(a) imply $M_{wd} > 0.66\,M_\odot$, unless the secondary is $M_2 < 0.045\,M_\odot$.

Thorstensen (2003) calculated the distance to EF Eri by combining information from parallax observations with a number of other pieces of information, including an estimate of the a-priori velocity distribution towards EF Eri. The constraint from parallax for EF Eri was weak, so the priors dominated his Bayesian calculation. This lead Thorstensen to give two distances, one with a velocity prior including a high-velocity population, leading to a large distance of 163 pc, and one without this population, leading to a short distance of 113 pc.

Inference on the likely distance can be made from the observed brightness and the derived temperature of the white dwarf. The effective temperature of the white dwarf is likely below 10000 K (Schwope et al. 2007, Szkody et al. 2010 submitted), although a temperature as high as 11000 K is not completely ruled out. The implied distances for the two assumed temperatures are illustrated in Fig. 5(b). A robust upper limit for the distance to the white dwarf assuming 11000 K is 168 pc, at 10000 K the upper limit distance is 144 pc, for an inclination of 60° the distance is 111 pc and 96 pc for degenerate/non-degenerate secondaries, respectively (assuming $T_{\rm eff} = 10000$ K). Fig. 5(a) also reveals that the orbital inclination has to be larger than 44° to not exceed Chandrasekhar's mass limit.

Taking $M_{wd} = 0.9\,M_\odot$ at face value the inclination becomes 60° or 66° (for a degenerate/non-degenerate secondary), in excellent agreement with optical/UV light curve modeling.

In Fig. 6 we show an updated version of the ultraviolet/optical/IR spectral energy distribution of the binary (update with respect to Schwope et al. 2007). Apart from an optical low-state spectrum adapted from Harrison et al. (2004) it comprises GALEX photometry at orbital minimum and maximum (Szkody et al. 2006), XMM-OM optical and visual photometry (Schwope et al. 2007, UV red dots), orbital phase-resolved infra-red photometry in JHK (Harrison et al. 2003) and SPITZER infrared data from Brinkworth et al. (2007) (for a Spitzer spectrum and a further discussion of the mid-IR spectral energy distribution see Hoard et al. 2007, data points shown as blue dots). The green lines illustrate the combined white-dwarf/accretion spot model adapted from Schwope et al. (2007).

Extrapolation of the white-dwarf/accretion-spot model into the infrared spectral regime shows that the most sensitive limits on the secondary can likely be derived from the J-band. The H and K bands have a much stronger dust component (Hoard et al. 2007).

The low-state infrared spectrum and variability is dominated by cyclotron emission at all phases (Campbell et al. 2008). The observed J-band minimum flux corrected for the contribution from the white dwarf thus represents an upper limit to the flux of the secondary, $f_{lim} = 9 \times 10^{-18}\,\mathrm{erg\,cm^{-2}\,s^{-1}\,\AA^{-1}}$ ($m_J = 18.87$). We searched Knigge's donor sequence for a secondary that does not violate the brightness limit for a given distance. This limit, the maximum possible absolute J-band magnitude of a donor, is shown in Fig. 5(c) as a function of the distance. In an erratum to the original publication Knigge (2007) gives $M_J = 13.2$ at the stellar/sub-stellar boundary. However, at period minimum his IR-sequences show wiggles due to switches in the model atmosphere grids used. At the stellar/sub-stellar boundary $M_J$ might be as bright as 12.6 if one interpolates the tabulated values with a smooth function. The region of uncertainty is indicate as a shaded box in Fig. 5(c). Fortunately, at $P = 81$ min we are away from this uncertain region on either side (stellar or sub-stellar). Hence from Fig. 5(c), the secondary has to be sub-stellar at any distance shorter than ∼140 (or ∼180 pc, if the boundary is at $M_J = 12.6$).

According to Knigge's donor sequence, a stellar secondary at the orbital period of 81 min has $M_J = 12.05$, and the combined flux of the white dwarf (including the accretion spot) plus such a secondary would not exceed the observed J-band minimum flux only at a distance of 230 pc (blue line in Fig. 6). Hence, if Knigge's sequence is applicable to EF Eri and if the observed J-band minimum flux is a valid proxy of the secondary's brightness, a stellar secondary in EF Eri can likely be ruled out.

A substellar Knigge-donor plus white dwarf will not exceed $m_J(min)$ at a distance of 135 pc (magenta line in Fig. 6). This value is rather close to the mentioned boundary of 140 pc and due to the non-monotonic behaviour of $M_J$ around period minimum in Knigge's tables.

## 3. Discussion and outlook

Using X-Shooter and exposing just 35 min on target we found good evidence for a massive white dwarf in EF Eri and a substellar secondary, establishing the object as a period bouncing CV at a distance of about 110 pc. The clues are based on line profile variations of the near-IR Ca II triplet which we found in accord with irradiation of the still unseen donor.

We note that even our *observed* radial velocity amplitude, $K'_2$, which we regard as originating from the irradiated secondary, is by $114\,\mathrm{km\,s^{-1}}$ larger than the H$\alpha$ velocity published by Howell et al. (2006) and interpreted as true $K_2$. This makes the discussion of the stellar masses presented there shaky. A Doppler tomographic analysis of the Ca II and the H$\alpha$ line of the X-Shooter data reveals the same velocities for the narrow emission line components. It appears likely that the H$\alpha$ line observed by Howell et al. (2006) has contributions from other locations in the binary system, e.g. the stream.

Our mass function was derived for two assumed secondaries from Knigge's CV donor sequence. Investigating period-bouncing, eclipsing, non-magnetic CVs Littlefair et al. (2008) found deviations between those predicted and the observed masses. The secondaries were found to be more massive than predicted. The same may apply here. We thus underline the importance of an improved parallax for EF Eri to better constrain the mass, hence radius of the white dwarf as well as an improved determination of the orbital inclination (e.g. via Ca II line flux measurements with higher signal-to-noise and full-phase coverage) to better constrain the mass ratio.

Nevertheless, the likely white dwarf mass found here of around $0.9\,M_\odot$ for $i = 60°$ is in accord with X-ray spectral fits (Cropper et al. 1998). Such a high mass would conform with the mean mass of magnetic white dwarfs, $0.93\,M_\odot$ (compared to $0.57\,M_\odot$ for non-magnetic white dwarfs), as found by Wickramasinghe & Ferrario (2005).

The direct discovery of the secondary in any short-period magnetic CV is an outstanding issue. Direct measurement or indirect evidence as presented here are very much important to compare the fraction of period-bouncing to pre-period minimum objects and thus test population synthesis predictions (Howell et al. 2001) and to test the CV donor sequence at the bottom end. Further X-Shooting of carefully selected polar systems is thus highly desirable.

*Acknowledgements.* We gratefully acknowledge constructive comments made by an anonymous referee. We thank the X-Shooter instrument team and the user support group at ESO for excellent support during preparation and reduction of the observations and for obtaining the spectra in service mode.